\documentclass[namedreferences]{solarphysics}
\usepackage[hyperref,linksfromyear,optionalrh]{spr-sola-addons} 
\usepackage{graphicx}                    
\usepackage{type1cm}
\usepackage{lmodern}
\usepackage{multirow}
\usepackage{url}
\usepackage{amsmath} 

\newcommand{\hei}{He\,{\sc i}~D3}
\newcommand{\fei}{Fe\,{\sc i}}
\newcommand{\ha}{H$\alpha$}
\newcommand{\kms}{km\,s$^{-1}$}
\newcommand{\heii}{304\,\AA}
\newcommand{\fexiv}{211\,\AA} 
\newcommand{\fexii}{195\,\AA} 
\newcommand{\degr}{$^{\circ}$}
\newcommand{\apj}{    {\it Astrophys. J.}}
\newcommand{\aap}{    {\it Astron. Astrophys.}}
\newcommand{\apjl}{   {\it Astrophys. J. Lett.}}
\newcommand{\solphys}{{\it Solar Phys.}}
\newcommand{\ssr}{    {\it Space Sci. Rev.}} 
\newcommand{\aaps}{   {\it Astron. Astrophys. Suppl.}}
\newcommand{\procspie}{   {\it Society of Photo-Optical Instrumentation Engineers (SPIE) Conference Series}}

\newcommand{\zap}{    {\it ZAp}} 
\newcommand{\mnras}{  {\it Mon. Not. Roy. Astron. Soc.}}

\begin{document}
\begin{article}
  \begin{opening}

    \title{Spectral Characteristics of the \hei\ Line \\ in a Quiescent Prominence Observed by THEMIS}

    \author[ addressref={aff1}, corref, email={koza@astro.sk} ]{\inits{J.}\fnm{J\'{u}lius}~\lnm{Koza}\orcid{0000-0002-7444-7046}}
    \author[ addressref={aff1} ]{\inits{J.}\fnm{J\'{a}n}~\lnm{Ryb\'{a}k}\orcid{0000-0003-3128-8396}}
    \author[ addressref={aff1} ]{\inits{P.}\fnm{Peter}~\lnm{G\"{o}m\"{o}ry}\orcid{0000-0002-0473-4103}}
    \author[ addressref={aff1} ]{\inits{M.}\fnm{Mat\'{u}\v{s}}~\lnm{Koz\'{a}k}}
    \author[ addressref={aff2,aff3} ]{\inits{A.}\fnm{Arturo}~\lnm{L\'{o}pez Ariste}}

    %
    \runningauthor{J.\,Koza {\it et al.}}
    \runningtitle{Spectral Characteristics of \hei\ in a Quiescent Prominence}

    \address[id=aff1]{Astronomical Institute, Slovak Academy of Sciences, Tatransk\'{a} Lomnica, Slovakia}
    \address[id=aff2]{Universit\'{e} de Toulouse, UPS-OMP, Institut de Recherche en Astrophysique et Plan\'{e}tologie, Toulouse, France}
    \address[id=aff3]{CNRS, Institut de Recherche en Astrophysique et Plan\'{e}tologie, 14 Avenue Edouard Belin, F-31400 Toulouse, France}
    \begin{abstract}
      We analyze the observations of a quiescent prominence acquired
      by the {\it T\'{e}l\'{e}scope Heliographique pour l'\'{E}tude du
        Magnetisme et des Instabilit\'{e}s Solaires} (THEMIS) in the
      He\,{\sc i}~5876\,\AA\ (\hei) multiplet aiming to measure the
      spectral characteristics of the \hei\ profiles and to find for
      them an adequate fitting model.
The component characteristics of the
\hei\ Stokes~{\it I} profiles
      are measured by the fitting system approximating them with a
      double Gaussian.
      This model yields an \hei\ component peak intensity ratio of
      $5.5\pm0.4$, which differs from the value of 8 expected in the
      optically thin limit. Most of the measured Doppler velocities
      lie in the interval $\pm5$\,\kms, with a standard deviation of
      $\pm1.7$\,\kms\ around the peak value of $0.4$\,\kms. The wide
      distribution of the full-width at half maximum has two maxima at
      0.25\,\AA\ and 0.30\,\AA\ for the \hei\ blue component and two
      maxima at 0.22\,\AA\ and 0.31\,\AA\ for the red
      component. The width ratio of the components is
      $1.04\pm0.18$. We show that the double-Gaussian model
      systematically underestimates the blue wing intensities. To
      solve this problem, we invoke a two-temperature multi-Gaussian
      model, consisting of two double-Gaussians, which provides a
      better representation of \hei\ that is free of the wing
      intensity deficit. This model suggests temperatures of 11.5\,kK
      and 91\,kK, respectively, for the cool and the hot component of
      the target prominence. The cool and hot components of a typical
      \hei\ profile have component peak intensity ratios of 6.6 and 8,
      implying a prominence geometrical width of 17\,Mm and an optical
      thickness of 0.3 for the cool component, while the optical
      thickness of the hot component is negligible.
      These prominence parameters seem to be realistic, suggesting the
      physical adequacy of the multi-Gaussian model with important
      implications for interpreting \hei\ spectropolarimetry by
      current inversion codes.

    \end{abstract}
    %

  \end{opening}

  \section{Introduction}


  The He\,{\sc i} multiplet at 5876\,\AA, called the D3 line
  (hereafter \hei), is one of the prime diagnostics used in
  ground-based observations of solar prominences
  \citep{LopezAriste2015}. The multiplet consists of six
  transitions. In current solar spectropolarimetric observations, a
  combined emission of five transitions is resolved as its stronger
  blue component, while the sixth transition makes the weaker red
  component. Their spectral separation is 343.3\,m\AA, giving the line
  its characteristic double-lobed profile. In the limiting case of
  negligible optical thickness in both components and the natural
  excitation of the triplet 3\,$^3$D term, the blue-to-red ratio of
  the component peak intensities is 8:1
  \citep{LandiDeglInnocenti1982}.


  The optical thinness of prominence plasma in \hei\ has been
  documented convincingly by i) the semi-empirical modeling of
  \hei\ profiles
  \citep{Landmanetal1977,Fontenla1979,Landman1981,Lietal2000}, ii) the
  non-LTE radiative modeling of
  \hei\ \citep{LabrosseandGouttebroze2001,LabrosseandGouttebroze2004,Leger2008,LegerandPaletou2009},
  and iii) the obvious observational fact of the on-disk transparency
  of filaments in \hei. The non-local thermal equilibrium (non-LTE) 2D
  multi-thread modeling by \citet{Leger2008} and
  \citet{LegerandPaletou2009} showed that under realistic conditions
  the prominence plasma never becomes optically thick for
  \hei. Remarkably, these studies also showed that an optically thin
  structure, composed of several optically thin threads, emits the
  \hei\ profiles with component peak intensity ratios lower than
  8. (We recommend a reader of \citet{LegerandPaletou2009} to change
  their Figure~9 for Figure~7.21 in \citet{Leger2008}.) This
  compatibility of the optical thinness with the ratio of lower than 8
  settles down earlier indications of non-negligible plasma optical
  thickness in \hei\ expressed in
  \citet{HouseandSmart1982,Athayetal1983a,LopezAristeandCasini2002,Casinietal2003};
  and \citet{WiehrandBianda2003} in the context of the observed ratios
  ranging from 6 to 7.6.
  

  The \hei\ emission has been used mainly in spectropolarimetry of
  prominences aiming to measure prominence magnetic fields. The
  pioneering work in this field by \citet{HarveyTandbergHanssen1968}
  was later followed by the series of studies by J.\,L.\,Leroy,
  V.\,Bommier, and S.\,Sahal-Br\'{e}chot \citep{Leroy1989}. The
  successful measurements made by these authors confirmed that the
  magnetic fields are essentially horizontal and twisted, with average
  field strengths of 10 -- 20\,G. Horizontal fields were detected in
  the majority of nearly 200 prominences observed from 2012 to 2015
  \citep{Levensetal2016a}. However, this study reported non-horizontal
  fields in a jet-like structure in an eruptive prominence, with
  inclinations tilted to 50\degr\ from the vertical. A recent
  multi-instrumental study involving \hei\ by \citet{Levensetal2016b}
  reported finding a horizontal magnetic field in the legs of a
  tornado-like prominence. \citet{Schmiederetal2013} also found mostly
  horizontal magnetic fields by \hei\ in a prominence pillar showing
  signatures of magnetosonic waves propagating transverse to the
  magnetic field with a velocity of about 10\,\kms, a period of about
  300\,s, and a wavelength of around 2\,000\,km. The low
  signal-to-noise ratio of the \hei\ polarization signatures was
  interpreted in \citet{Schmiederetal2014} as a possible manifestation
  of a turbulent field superimposed on the macroscopic horizontal
  component of the prominence magnetic field. The results of these
  recent studies made use of \hei\ spectropolarimetry obtained by the
  {\it T\'{e}l\'{e}scope Heliographique pour l'\'{E}tude du Magnetisme
    et des Instabilit\'{e}s Solaires} (THEMIS) solar telescope.


  Current models of prominences or filaments present them as
  aggregates of thin threads that are weakly or highly twisted into
  flux ropes with the effective volume-filling factor of radiating
  threads on the order of $10^{-1}-10^{-3}$
  \citep{Mackayetal2010,Engvold2015}. These models stem from
  high-resolution observations showing their fine-thread and knotty
  composition with thread widths down to the resolution limit of
  $\approx150$\,km
  \citep{Linetal2005,Linetal2007,Linetal2008,HeinzelAnzer2006,Gunaretal2007,Gunaretal2008,Kuckeinetal2014,Freedetal2016}. A
  significant advance in modeling the fine structure of prominences
  and filaments has been achieved in the studies by
  \citet{GunarMackay2015a,GunarMackay2015b}, which for the first time
  present the whole-prominence and filament fine-structure model and
  their synthetic \ha\ images in 3D. This study even presented the
  temporal evolution of prominences and filaments in response to
  changes in the underlying photospheric magnetic flux distribution,
  and it visualized their \ha\ appearance as it evolved in time. In an
  earlier study, \citet{Gunaretal2012} statistically tested the 2D
  multi-thread model of a quiescent prominence comparing the
  integrated intensity, the full-width at half maximum (hereafter FWHM
  or just the width), and the Doppler velocity of the observed and
  synthetic \ha\ profiles computed by the model. Of course, upcoming
  or future radiative transfer codes \citep[see][]{Stepanetal2015} and
  multi-thread models of prominences or filaments should allow the
  comparison of observed and synthetic \hei\ Stokes
  profiles. \citet{Leger2008} and \citet{LegerandPaletou2009} have
  presented the results of 2D non-LTE modeling of the \hei\ line
  taking the atomic fine structure of helium into account.


  The aims of this study are twofold: first, to produce solid
  statistics of spectral characteristics of the blue and red component
  of \hei\ in a form that allows comparisons with future
  \hei\ modeling in the fashion applied in \citet{Gunaretal2012} for
  \ha; second, to suggest an adequate fitting model for the
  \hei\ Stokes~{\it I} profile in accord with contemporary views on
  prominence thermodynamic structure. The definition of these aims is
  motivated by a lack of pertinent studies.

  \section{Observations}  

\begin{figure}
      \centerline{\includegraphics{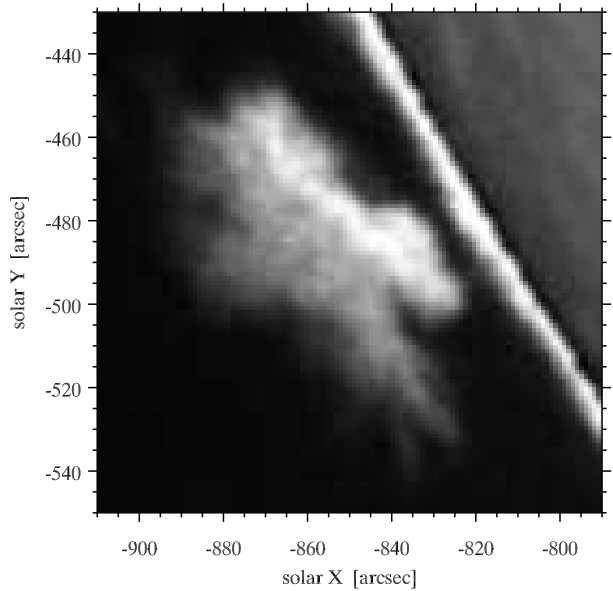}}
      \caption{The cutout from the \ha\ broad-band full-disk image of
        the target prominence at the position angle of
        117\degr\ obtained at the Kanzelh\"{o}he Solar Observatory on
        2 August 2014 at 14:12:39\,UT. The THEMIS observation started
        at 14:59:02\,UT. Different intensity scaling is applied to the
        disk and off-limb domain of the image.}
      \label{fig1}
  \end{figure}

  A tree-like quiescent prominence occurring on 2 August 2014 at the
  east solar limb at the position angle of
  117\degr\ (Figure~\ref{fig1}) was selected as a target for the
  double-beam spectropolarimetry in the \hei\ line by THEMIS operated
  in the MulTi Raies mode at Observatorio del Teide
  \citep{LopezAristeetal2000}. The prominence area in the plane of the
  sky, measured in the \ha\ broad-band image (Figure~\ref{fig1}) taken
  at the Kanzelh\"{o}he Solar Observatory
  \citep[KSO,][]{Poetzietal2015}, is 1483\,Mm$^2$. After subtracting
  the off-limb aureole that is due to the scattered photospheric
  light, the prominence relative intensity in the KSO 0.7\,\AA-wide
  \ha\ Lyot filter with respect to the disk center is within the range
  from 7\% to 27\% with a median of 17\%, corresponding to a
  radiance of 1.0\,W\,cm$^{-2}$\,sr$^{-1}$, and the prominence
  \ha\ radiant intensity is about $3.9\times10^{22}$\,W\,sr$^{-1}$.

  \subsection{THEMIS Spectropolarimetry}
  \label{themis}  

\begin{figure}
    \centerline{
      \includegraphics[bb= 8 0 155 114, width=0.49\textwidth]{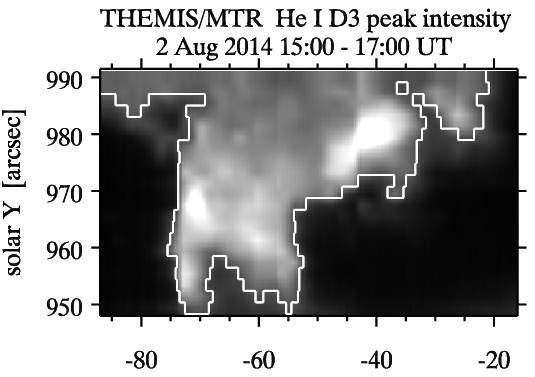}
      \hspace*{-0.05\textwidth}
      \includegraphics[bb= 0 0 147 118, width=0.49\textwidth]{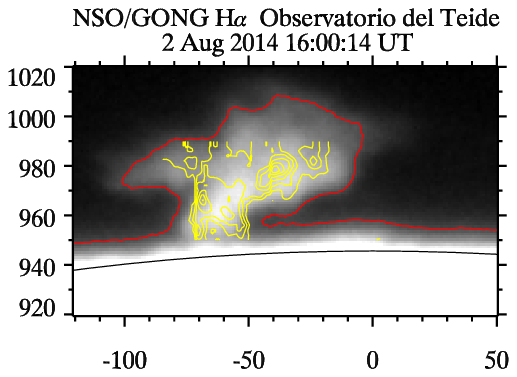}
    }
    \vspace{-0.03\textwidth}
    \centerline{
      \includegraphics[bb= 8 0 155 117, width=0.49\textwidth]{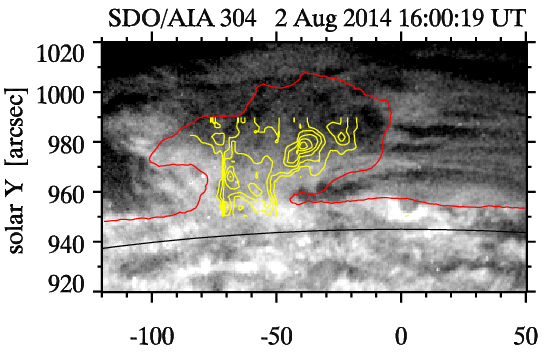}
      \hspace*{-0.05\textwidth}
      \includegraphics[bb= 0 0 147 117, width=0.49\textwidth]{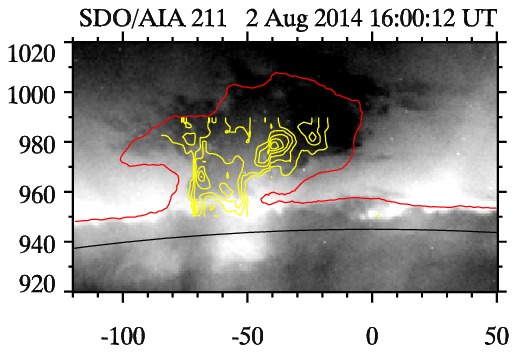}
    }
    \vspace{-0.01\textwidth}
    \centerline{
      \includegraphics[bb= 8 0 155 120, width=0.49\textwidth]{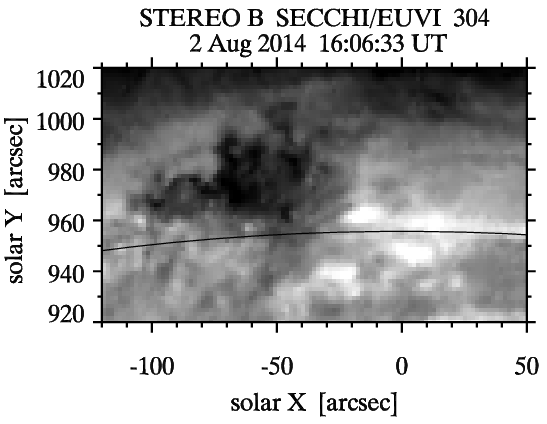}
      \hspace*{-0.05\textwidth}
      \includegraphics[bb= 0 0 147 120, width=0.49\textwidth]{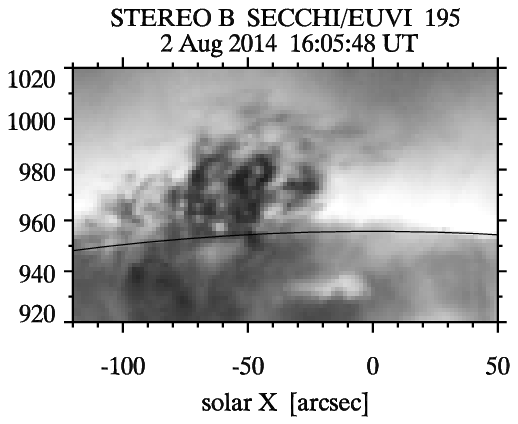}
    }
    \caption{{\it Top left:} Slit-reconstructed map of the \hei\ peak
      intensity of the target prominence observed by THEMIS. The
      scanning ran from about 15:00\,UT to 17:00\,UT. The solar limb
      is out of the image below its lower border. Note the different
      fields of view of this and the following panels. The {\it
        contour} outlines the prominence mask applied to extract the
      data shown in Figures~\ref{fig6} -- \ref{fig8}. {\it Top right:}
      The broad-band \ha\ image of the prominence taken by the
      NSO/GONG network station at the Observatorio del Teide
      overplotted. The {\it red contour} marks the selected
      \ha\ intensity level. The {\it yellow contours} show the
      coaligned \hei\ peak intensity of the prominence in the {\it
        left panel.} The {\it black arc} indicates the east
      photospheric limb. {\it Middle:} The SDO/AIA images of the
      prominence in the \heii\ ({\it left}) and \fexiv\ ({\it right})
      passbands, respectively. The {\it contours} and the {\it black
        arcs} have the same meaning as in the {\it top right
        panel}. {\it Bottom:} STEREO\,B SECCHI/EUVI images of the
      prominence in the \heii\ ({\it left}) and \fexii\ ({\it right})
      passbands, respectively. The {\it black arc} indicates the west
      photospheric limb from the spacecraft viewpoint. The NSO/GONG
      \ha, SDO/AIA, and SECCHI/EUVI images are taken approximately in
      the middle of scanning. The temporal evolution of the prominence
      in the NSO/GONG \ha\ and in the SDO/AIA passbands during the
      scanning is shown in the movie that is available in the online
      edition. The movie is assembled from 120 frames with a temporal
      resolution of 1\,min.}
    \label{fig2}
  \end{figure}

\begin{figure}
    \centerline{
      \hspace*{0.04\textwidth}
      \includegraphics[width=0.55\textwidth]{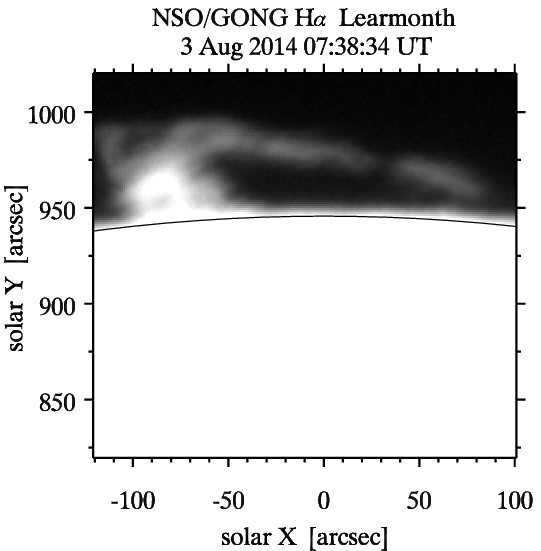}
      \hspace*{-0.07\textwidth}
      \includegraphics[width=0.55\textwidth]{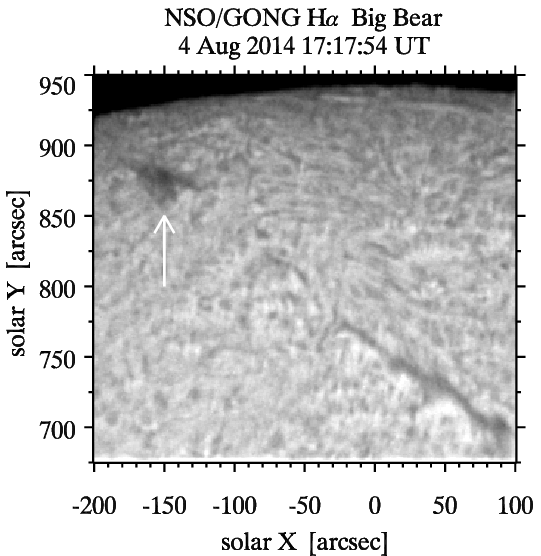}
    }
    \caption{The broad-band \ha\ images of the target prominence on 3
      August 2014 ({\it left}) and its probable on-disk filamentary
      counterpart on 4 August 2014 ({\it right}) marked by the {\it
        arrow}, taken at Learmonth and Big Bear NSO/GONG network
      stations, respectively. The {\it black arc} indicates the east
      photospheric limb.}
    \label{fig3}
  \end{figure}

  The slit of the THEMIS spectrograph was opened to one arcsecond and
  placed parallel to the limb. The setup involved the Semel mask, {\it
    i.e.} a grid mask with three rectangular windows placed at the F1
  focus of the telescope before the polarization analyzer
  \citep{Semel1980,BriandCeppatelli2002}. A rationale for its use and
  the necessity of 2D scanning in directions parallel ({\it X}) and
  perpendicular ({\it Y}) to the slit were clarified in
  \citet{SainzDaldaandLopezAriste2007} and
  \citet{Schmiederetal2013}. The 2D scanning of the target prominence
  started close to the limb at 14:59:02\,UT and progressed radially
  away from the limb untill about 17:00\,UT. It involved one step in
  the {\it X} direction with the size of 15\,arcsec followed by a step
  in the {\it Y} direction. In total, 25 steps separated by 2\,arcsec
  were ordered in the {\it Y} direction. This provided a field of view
  of 88\,arcsec\,$\times 50$\,arcsec, and the top left panel of
  Figure~\ref{fig2} shows the effective cutout covering the whole
  target prominence. The axes keep the ({\it X, Y}) orientation
  defined above. Unlike Figure~\ref{fig1}, in Figures~\ref{fig2},
  \ref{fig3}, \ref{fig6}, and \ref{fig7}, the solar disk and the
  target prominence are rotated about 117\degr\ clockwise (the THEMIS
  images; the image provided by the {\it Global Oscillation Network
    Group} \ha\ network monitor, operated by the {\it National Solar
    Observatory} (NSO/GONG); and the images taken by the {\it
    Atmospheric Imaging Assembly} onboard the {\it Solar Dynamic
    Observatory} (SDO/AIA)) or counterclockwise (the images taken by
  the {\it Solar TErrestrial RElations Observatory B} (STEREO\,B)) but
  keeping the same heliocentric coordinate system as in
  Figure~\ref{fig1}. The STEREO\,B images were then flipped over
  around the {\it Y} axis. The white contour in the top left panel of
  Figure~\ref{fig2} outlines the prominence mask defined in
  Section~\ref{data}.

  One particular THEMIS spectrum at a given slit position is taken
  with the exposure time of 2\,s, and overall, the observing procedure
  is the same as in \citet{Schmiederetal2013,Schmiederetal2014} and
  \citet{Levensetal2016a,Levensetal2016b}. We emphasize that in our
  case a modulation cycle of six images, spanning the three
  polarization states with either positive or negative signs, is
  repeated ten times to increase the signal-to-noise ratio. Therefore
  the 2D scanning lasts two hours.  A typical acquisition time of four
  Stokes parameters at one slit position is 144\,s, which includes the
  switch time of retarders and a dead time of the system duty cycle.


  The spatial and spectral sampling of the obtained spectra is
  0.227\,arcsec\,px$^{-1}$ and 11.6\,m\AA\,px$^{-1}$,
  respectively. The latter value is determined by the photospheric
  lines \fei~5873.218\,\AA\ and \fei~5877.797\,\AA\ that are seen in
  the spectra in the scattered photospheric light in the background
  (Figure~\ref{fig4}). The dispersion of $11.6$\,m\AA\,px$^{-1}$ is
  used in calculating the spectral characteristics presented in
  Section~\ref{results} and in Figures~\ref{fig6} -- \ref{fig7}.


  Comparing the average \fei\ line profiles with their counterparts
  extracted from the spectral atlas of solar disk-center intensity
  \citep{Neckel99} and convolved with a Gaussian, we estimate the
  instrumental smearing of the THEMIS spectra. We find that the
  smearing profile can be approximated by a Gaussian with FWHM =
  71\,m\AA.

  \subsection{Morphology, Dynamics, and Evolution of the Target Prominence}

  To place the THEMIS observations in context, we make use of data
  provided by NSO/GONG \citep{Harveyetal2011}, SDO/AIA
  \citep{Lemenetal2012,Pesnelletal2012}, and the {\it Extreme
    Ultraviolet Imager} (EUVI), which is a part of the {\it Sun-Earth
    Connection Coronal and Heliospheric Investigation} (SECCHI)
  instrument suite onboard the STEREO\,B spacecraft
  \citep{Wuelseretal2004,Wuelseretal2007,Howardetal2008}. Figure~\ref{fig2}
  and the movie available in the online edition show the
  slit-reconstructed map of the \hei\ peak intensity of the target
  prominence coaligned with the context images taken in the broad-band
  \ha\ filter of the NSO/GONG monitor at Observatorio del Teide and in
  the AIA \heii\ and \fexiv\ passbands during the THEMIS scanning
  overlaid with selected \hei\ peak intensity contours (yellow) and
  the \ha\ intensity contour (red). The coalignment of the \hei\ peak
  intensity image and the \ha\ image is approximate relying on
  morphological similarity. The coalignment of the \ha\ and AIA images
  is exact; it takes advantage of full-disk images and availability of
  all necessary data. The {\it X}-axis represents the position along
  the slit of the THEMIS spectrograph, and the {\it Y}-axis
  corresponds to the radial scanning direction. To highlight the
  structure of the target prominence, the top left panel shows a
  smaller field of view than the other panels. The \hei\ image
  suggests that the tree-like prominence consists of a central
  vertical pillar spreading out with height on both sides. The
  appearance of the prominence in \ha\ (top right panel) corresponds
  to the \hei\ peak intensity contours. The middle right panel shows
  the prominence in the AIA \fexiv\ passband as a dark structure
  against the bright background, absorbing the background coronal
  emission. Its ``silhouette'' agrees approximately with the shapes
  outlined by the \hei\ and \ha\ contours. However, the prominence
  appears very differently in the AIA \heii\ passband, which is shown
  in the middle left panel. It clearly illustrates the large areal
  extent and the horizontal arcades that stretch to the right, which
  are not visible in the previous panels.


  Quiescent prominences are often characterized as sheets of plasma
  standing vertically above the polarity inversion line and showing
  fine horizontal or vertical threads \citep{OrozcoSuarezetal2014}.
  The attained spatial resolution in the THEMIS and the NSO/GONG
  \ha\ observations prevents us from resolving any prominence fine
  structures. However, we do see signatures of strands in the AIA
  \fexiv\ and particularly in the \heii\ images. The bottom panels of
  the movie display plasma flows along the arcades, stretching to the
  right, that are generally parallel to the solar limb, similar to
  those described in \citet{Chaeetal2008} and
  \citet{OrozcoSuarezetal2014}. On the other side of the prominence
  body, one can see clumps of plasma flowing mostly toward the limb
  only in the AIA \heii\ passband.


  To gain a comprehensive picture of the target prominence, we also
  inspect images taken by the STEREO\,A and B spacecraft on 2 August
  2014 from 15:00\,UT to 17:00\,UT. While for STEREO\,A the prominence
  is invisible beyond the limb, the STEREO\,B images show it clearly
  at the west limb in all four passbands of the SECCHI/EUVI
  imager. Since the separation angle of the spacecraft with Earth is
  162\degr, the SECCHI/EUVI \heii\ and \fexii\ images of the
  prominence in the bottom panels of Figure~\ref{fig2} provide almost
  rear views of its dark central pillar and arcades stretching to the
  right, resembling its AIA \heii\ counterpart in the middle left
  panel. Different positions of the photospheric limbs in the SDO/AIA
  and STEREO\,B images are due to different solar disk diameters as
  seen from the vantage points of the spacecraft at heliocentric
  distances of about 1.015\,AU and 1.003\,AU, respectively.


  The post-observation NSO/GONG \ha\ image from 3 August 2014 in the
  left panel of Figure~\ref{fig3} shows that the prominence is still
  bright with a long and prominent arcade. The arrow in the NSO/GONG
  image from the following day (right panel) identifies its probable
  on-disk filamentary counterpart, which occupies a quiet-Sun area far
  from active regions. The SDO/HMI magnetogram (not shown here)
  suggests only an enhanced network in that area. Therefore, we
  classify the prominence as of quiescent type. It seems to be an
  isolated segment of the filament stretching diagonally from the
  lower right corner of the image. On that day, the marked structure
  disappears quickly.

  \section{Data Reduction} 
  \label{data}

\begin{figure}
    \centerline{\includegraphics[width=0.91\textwidth]{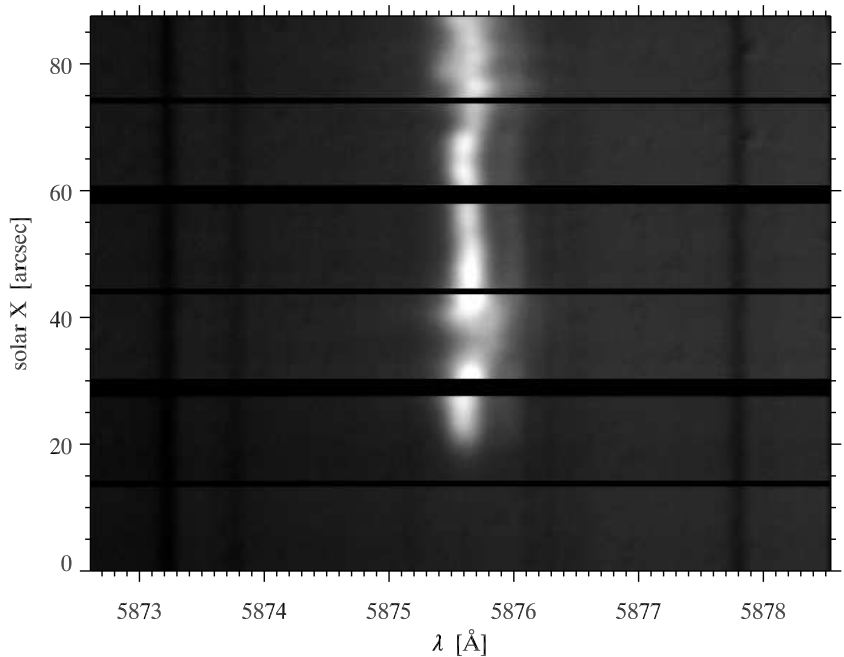}}
    \caption{An example of the THEMIS spectrum showing
      \hei\ Stokes~{\it I}, taken at the end of scanning when the slit
      is at $Y = 990$\,arcsec (the top left panel in
      Figure~\ref{fig2}). The stronger blue component of \hei\ and its
      weaker red component are well resolved in the {\it X} span from
      45\,arcsec to 75\,arcsec. The absorption features at about
      5873.2\,\AA\ and 5877.8\,\AA, seen in the scattered continuum
      light, are the photospheric \fei\ lines. The {\it black
        horizontal strips} are due to the Semel mask. The {\it X}-axis
      is parallel to the limb.}
    \label{fig4}
  \end{figure}

  The primary reduction is the same as applied in
  \citet{Schmiederetal2013,Schmiederetal2014} and
  \citet{Levensetal2016a,Levensetal2016b}, and it involves data from all
  ten modulation cycles. The raw THEMIS spectra are reduced using the
  IDL package DeepStokes, whose main characteristics are outlined in
  \citet{LopezAristeetal2009}. The data reduction includes geometry
  corrections of inclination and curvature of spectral lines, dark
  current and bias subtraction, flat-field correction, and a careful
  handling of the polarization signals. Since the reduction yields
  Stokes~{\it Q, U, V} profiles with very small amplitudes, typically
  0.2 -- 0.4\% of the Stokes~{\it I} peak intensity, we repeat the
  reduction taking data only from five modulation cycles as in
  \citet{Schmiederetal2013,Schmiederetal2014} and
  \citet{Levensetal2016a,Levensetal2016b}, but without
  improvement. Therefore, we aim the analysis only at the Stokes~{\it
    I} profiles. To eliminate the large temporal smearing, we repeat
  the DeepStokes reduction again, but taking data at each slit position
  only from the fifth, {\it i.e.} only one modulation cycle, lasting
  about 14\,s. Figure~\ref{fig4} shows an example of the one-cycle
  Stokes~{\it I} spectra.

  To extract only prominence-relevant data, we construct the
  prominence mask as a vector of pixel subscripts with the Stokes~{\it
    I} peak intensity $I_{\rm max}$ greater than or equal to the {\it
    ad-hoc} threshold value $0.2I_{\rm gmax}$, where $I_{\rm gmax}$ is
  the global maximum of the peak intensity over the whole
  prominence. The condition $I_{\rm max} \geq 0.2I_{\rm gmax}$ is
  fulfilled at the 2983 pixel positions defined in the top left panel
  of Figure~\ref{fig2} by the white contour representing the
  prominence in Figures~\ref{fig6} and \ref{fig7}. The data gaps due
  to the Semel mask (Figure~\ref{fig4}) are filled in by linear
  interpolation.

  \section{Double-Gaussian Fit}
  \label{Double-Gaussian-Fit}        


  Prominence models are often constructed and validated by comparing
  spectral characteristics of observed and synthetic line profiles
  \citep[\textit{e.g.}][]{Gunaretal2010,Schwartzetal2015}. The study
  by \citet{Gunaretal2012} made use of this approach in modeling the
  fine structure of a quiescent prominence by statistically comparing
  the Doppler velocity, the integrated intensity, and the FWHM of
  observed and synthetic \ha\ line profiles computed for an array of
  multi-thread prominence models. To provide solid statistics of
  spectral characteristics for the \hei\ Stokes~{\it I} profiles, we
  perform a double-Gaussian fitting of their double-lobe structure
  resolved in the THEMIS spectra (Figure~\ref{fig4}). The background
  intensity is approximated by a linear fit. The choice of this
  conservative fitting model is justified by the studies of, {\it
    e.g.}, \citet{HouseandSmart1982} and
  \citet{LandiDeglInnocenti1982}, who claimed that the double Gaussian
  is an adequate representation of the \hei\ double-lobed profile.

  \begin{figure}
    \centerline{
      \includegraphics{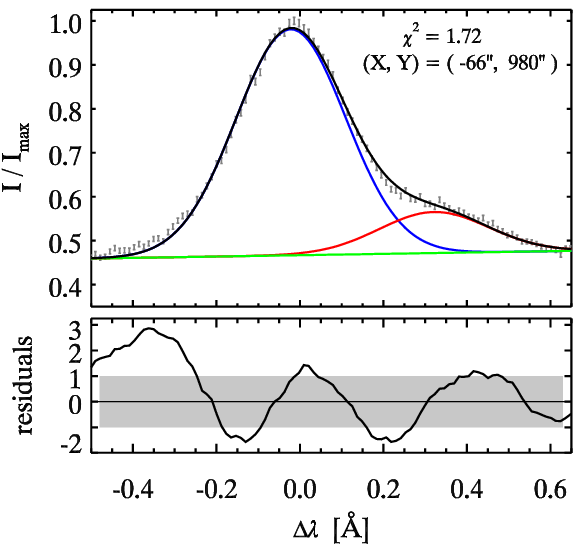}
      \includegraphics{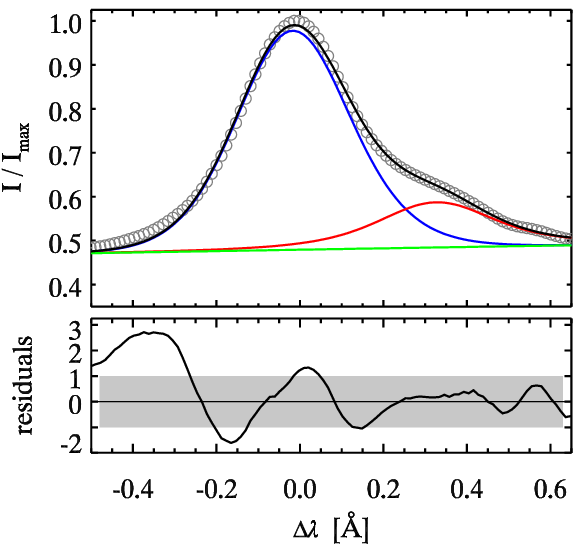}
    }
    \caption{{\it Left:} An example of a typical double-Gaussian fit
      ({\it black}) of an observed \hei\ Stokes~{\it I} profile ({\it
        gray error bars}) with a $\chi^2$ value of 1.72. The blue and
      red Gaussian components and the linear background are
      represented by the {\it blue}, {\it red}, and {\it green
        lines}. {\it Right:} The averages of all 2983 profile fits
      ({\it black}), observations ({\it gray circles}), the blue and
      red Gaussian components, and the linear backgrounds. The {\it
        bottom subpanels} display the best-fit residuals expressed in
      the one-sigma unit shown as the shaded rectangle.}
    \label{fig5}
  \end{figure}


  For the background and double-Gaussian fitting we use the SolarSoft
  function \texttt{mpfitfun.pro} calling the engine \texttt{mpfit.pro}
  \citep{Markwardt2009,More1978,MoreWright1993}. The latter performs
  Levenberg-Marquardt least-squares minimization of the sum of the
  weighted squared differences between the data $I_i\;(i = 1 \dots N)$
  and the user-supplied model function $F_i$. The relevant merit
  function is defined as $\chi^2 = \frac{1}{\rm N}\sum_{i=1}^{\rm
    N}(I_i - F_i)^2/\sigma_i^2$, where the weights $\sigma_i$
  represent one-sigma uncertainties of the data. Assuming Poisson
  statistics for propagation of the uncertainties, we estimate these
  as the standard deviation of the background intensity variations
  $\sigma_{\rm back}$ outside the spectral lines and rescale them
  using the formula $\sigma_i = \sigma_{\rm back} \sqrt{I_i / \left<
    I_{\rm back} \right>}$, where $\sigma_i$ and $I_i$ are the
  uncertainty and intensity at the particular wavelength and $\left<
  I_{\rm back} \right>$ is the average background intensity outside
  the lines. The definition of the merit function implies that the
  $\chi^2$ values of good fits are about one, meaning that the model
  is within the one-sigma uncertainties of the data.


  We approximate the observed \hei\ Stokes~{\it I} profiles with a
  model of two Gaussians superimposed on the first-order polynomial
  representing the background intensity. The model has seven free
  parameters: the peak intensities $I_{\rm blue,\, red}$, the
  FWHM$_{\rm blue,\, red}$, the spectral position of the blue
  component peak $\lambda_{\rm blue}$, and the two coefficients of the
  polynomial. The spectral position of the red component peak is tied
  to $\lambda_{\rm blue}$ as $\lambda_{\rm red} = \lambda_{\rm blue} +
  343.3$\,m\AA. This coupling is kept in all models discussed
  below. We also check a six-parameter model with equal widths of the
  components. This model yields component peak intensity ratios that
  are different from the optically thin limit, suggesting radiative
  transfer effects. Therefore, we decide not to tie the widths and to
  adopt the seven-parameter model with the red component width as a
  free parameter.


  Figure~\ref{fig5} shows an example of the fit of a typical profile
  and an average of all 2983 profile fits in the top left and right
  panels, respectively. The corresponding fit parameters are listed in
  columns 1 and 2 of Table~\ref{tab1}. The bottom subpanels display
  the best-fit residuals $(I_i - F_i)/\sigma_i$, normalized by the
  uncertainties of observations $\sigma_i$. They show that the
  residuals of typical fits are mostly within the $1\sigma$ span. The
  largest residuals of about $3\sigma$ occur at $\Delta\lambda \approx
  -0.35$\,\AA, suggesting a systematic excess of the observed blue
  wing intensities over the Gaussian wing profile. We refer to this
  feature in the following as the blue wing excess.
  
  \section{Results of the Double-Gaussian Fitting}
  \label{results}

\begin{figure}
    \centerline{
      \includegraphics{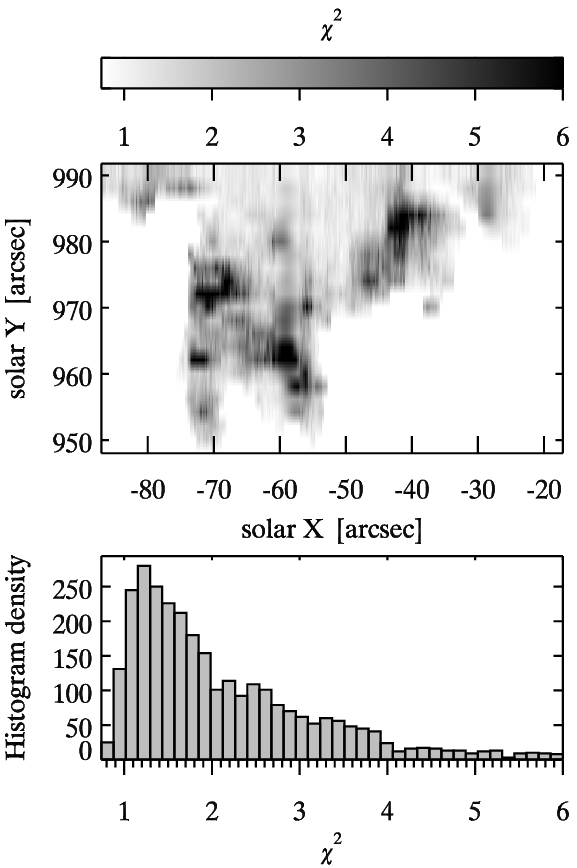}
      \includegraphics{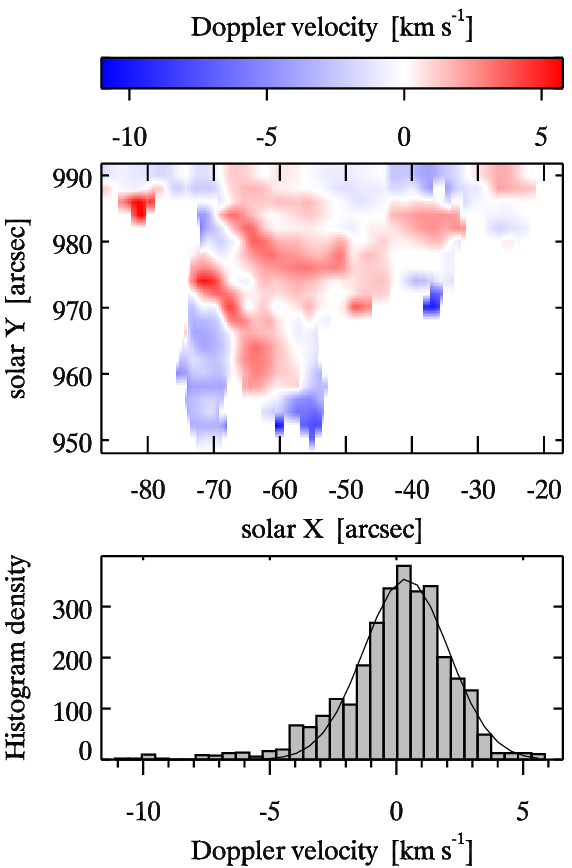}
    }
        \vspace{0.0125\textwidth}
    \centerline{
      \includegraphics{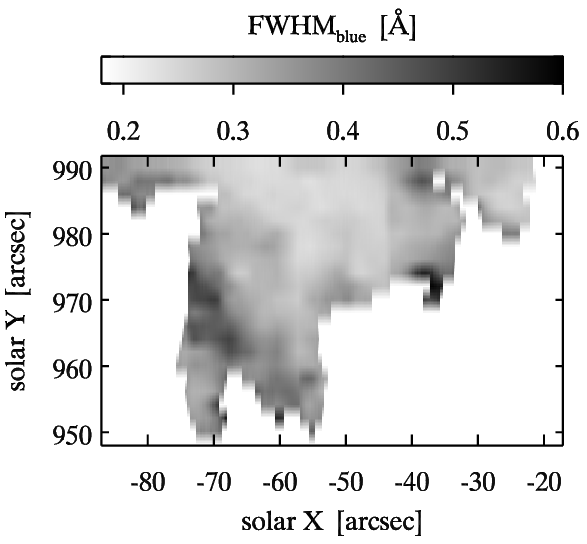}
      \includegraphics{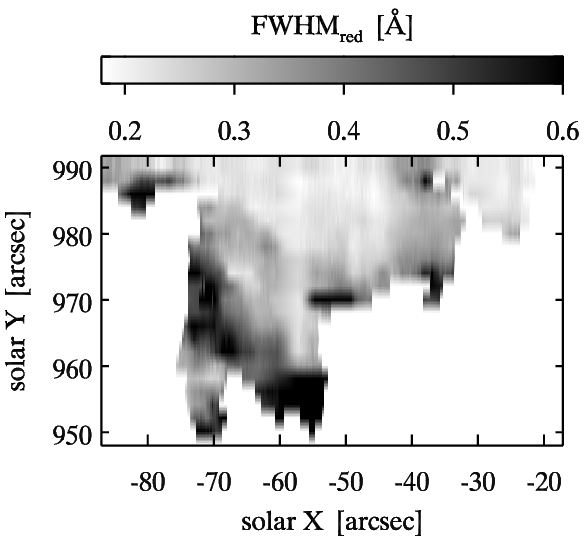}
    }
    \includegraphics{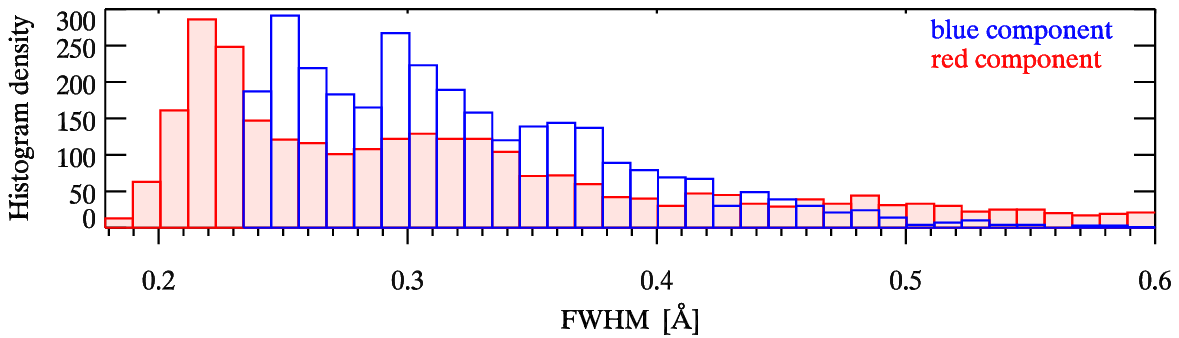}
    \caption{Maps and histograms of $\chi^2$ values and spectral
      characteristics of the \hei\ line observed in the target
      prominence. {\it Top left:} The $\chi^2$ values characterizing
      the best double-Gaussian fits. {\it Top right:} The Doppler
      velocity. The positive ({\it red}) is the redshift of the line
      center. The Gaussian fit of the histogram is indicated by the
      {\it thin line}. {\it Bottom:} FWHM of the blue ({\it left}) and
      the red ({\it right}) component of the \hei\ line.}
    \label{fig6}
  \end{figure}

  Figure~\ref{fig6} shows maps and histograms of $\chi^2$ values and
  spectral characteristics of the observed double-lobed Stokes~{\it I}
  profiles inferred by the double-Gaussian fitting of the one-cycle
  Stokes~{\it I} profiles. The $\chi^2$ distribution in the top left
  subpanel has a maximum and median at 1.2 and 1.8, respectively. Most
  of the measured Doppler velocities lie in the interval
  $\pm5$\,\kms, and their distribution shows a redshifted median and
  peak at 0.3\,\kms\ and $0.4\pm1.7$\,\kms, respectively. The latter
  value and its standard deviation result from the Gaussian fit of the
  distribution. An average of all 2983 pixel positions of blue
  component maxima is taken as a reference for the Doppler velocity.


  The bottom panels of Figure~\ref{fig6} show maps of spatial
  distributions and histograms of the widths of the blue and red
  components. The maps of FWHM$_{\rm blue}$ and FWHM$_{\rm red}$
  suggest a frequent occurrence of narrow profiles with component
  widths smaller than 0.3\,\AA\ in a compact area in the center of the
  upper part of the prominence body at $(X, Y) \approx (-55,
  980)$\,arcsec. The area coincides approximately with a large island
  of redshifts in the velocity map. The FWHM$_{\rm blue}$ histogram
  has not one global maximum, but suggests a double-peaked
  distribution with two local maxima at 0.25\,\AA\ and
  0.30\,\AA. Similarly, the distribution of FWHM$_{\rm red}$ has a
  prominent peak at 0.22\,\AA\ and a second peak at 0.31\,\AA. The
  medians of the FWHM$_{\rm blue}$ and FWHM$_{\rm red}$ histograms are
  at 0.31\,\AA\ and 0.29\,\AA, respectively. A comparison of the two
  histograms suggests an excess of red components narrower than
  0.23\,\AA\ and broader than 0.48\,\AA\ compared with the blue
  components.

\begin{figure}
    \centerline{
      \includegraphics{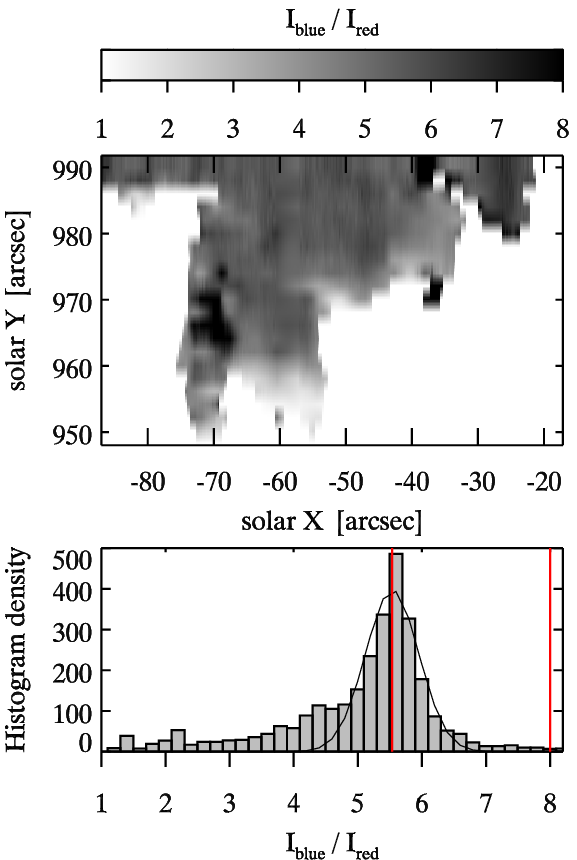}
      \includegraphics{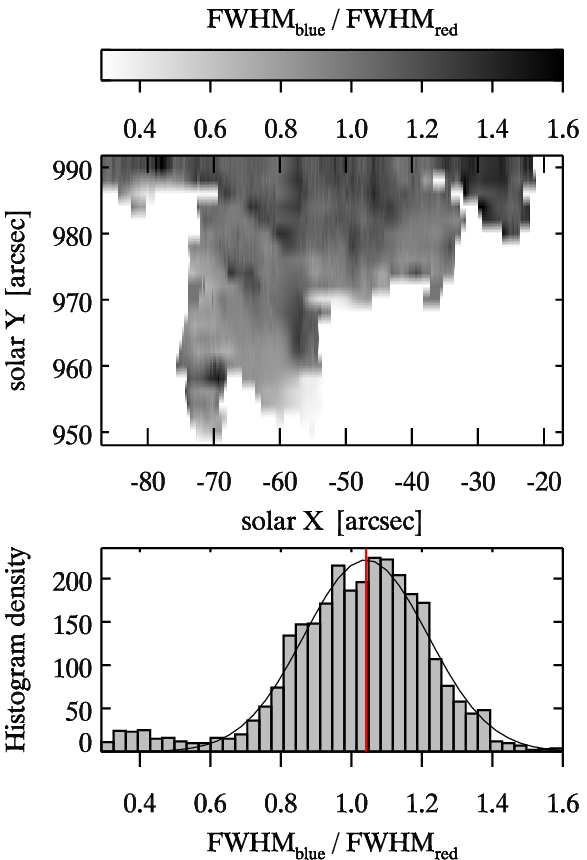}
    }
    \caption{Ratios of the peak intensities $I_{\rm blue}/I_{\rm red}$
      ({\it left}) and the FWHM$_{\rm blue}$/FWHM$_{\rm red}$ ({\it
        right}) of the blue and red components of the \hei\ line. The
      Gaussian fits of the histograms are indicated by the {\it thin
        lines}. The {\it red vertical lines} at 5.5 and 1.04 mark the
      peaks of the Gaussian fits. The {\it red line} at 8.0 indicates
      the value required by the optically thin limit.}
    \label{fig7}
  \end{figure}

  An indicator of the plasma optical thickness in \hei\ is the ratio
  of peak intensities and widths of the \hei\ double-lobed profile
  \citep{Leger2008}. The distribution of the peak intensity ratios in
  the lower left subpanel of Figure~\ref{fig7} has a median and peak
  at 5.4 and $5.5\pm0.4$, respectively.  The latter value and its
  standard deviation result from the Gaussian fit of the distribution,
  which shows an excess of profiles with lower ratios down to 2. The
  distribution of the width ratios in the lower right subpanel has a
  median and peak at 1.03 and $1.04\pm0.18$.
  

  We emphasize that the typical observed profile and its fit in the
  left panel of Figure~\ref{fig5} are chosen to be representative in
  the sense of typical values of $\chi^2 = 1.72$, a blue component
  width of 0.31\,\AA, and a component peak intensity ratio of
  5.5 (see also the corresponding entries in Tables~\ref{tab1} and
  \ref{tab2}).

\begin{figure}[!ht]
    \centerline{
      \includegraphics[width=0.5\textwidth]{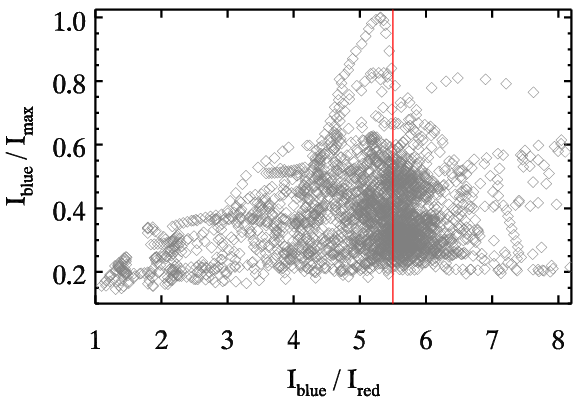}
      \includegraphics[width=0.5\textwidth]{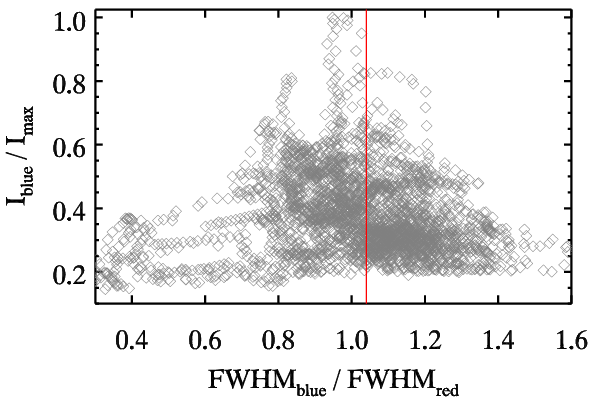}
    }
    \caption{Scatter plots between measured quantities for all pixels
      within the target prominence. The \hei\ blue component peak
      intensity normalized to its maximum as a function of the ratios
      of the peak intensities $I_{\rm blue}/I_{\rm red}$ ({\it left})
      and the FWHM$_{\rm blue}$/FWHM$_{\rm red}$ ({\it right}). The
      {\it red vertical lines} at 5.5 and 1.04 mark the values
      inferred from the histograms in Figure~\ref{fig7}.}
    \label{fig8}
  \end{figure}

  Figure~\ref{fig8} presents pixel-by-pixel comparisons of the blue
  component peak intensities $I_{\rm blue}$ with the peak intensity
  ratios (left panel) and the width ratios (right panel) in the form
  of scatter plots supplementing the histograms in
  Figure~\ref{fig7}. The plots suggest that while the weak profiles
  show a broad range of ratios, the strong profiles with high peak
  intensities $I_{\rm blue}$ have peak intensity ratios and width
  ratios of about $5.5\pm0.4$ and $1.04\pm0.18$, respectively. These
  are indicated by the red vertical lines. The shape of the scatter
  plots also suggests that the values $5.5\pm0.4$ and $1.04\pm0.18$
  inferred from the histograms in Figure~\ref{fig7} are representative
  of weak as well as strong profiles.

  \subsection{Effects of Temporal and Instrumental Smearing}
  \label{effect} 


  To examine the effect of temporal smearing on the spectral
  characteristics of the target prominence, we calculate them using
  the ten-cycle Stokes~{\it I} profiles with an effective integration
  time of 144\,s. Since this trial yields the same results as those
  presented in Figures~\ref{fig6} and \ref{fig7}, we conclude that the
  dynamics of the target quiescent prominence evolves on a timescale
  longer than two minutes at spatial scales larger than 2\,arcsec.


  The \hei\ profiles analyzed in this section are the convolutions of
  the original emission profiles with the instrumental profile of the
  THEMIS spectrograph, which can be approximated by a Gaussian with
  FWHM = 71\,m\AA\ (see Section~\ref{themis}). How much does this
  instrumental smearing bias the results? Could the smearing be
  responsible for the blue wing excess? To answer these questions, we
  deconvolve the typical \hei\ profile shown in the left panel of
  Figure~\ref{fig5} by applying the optimum filter
  \citep{BraultandWhite1971}. The deconvolved \hei\ profile is shown
  in the left panel of Figure~\ref{fig9} by the gray circles, and its
  fit is plotted with the black line. We apply the seven-parameter
  double-Gaussian model in fitting the deconvolved profile. Columns 1
  and 3 of Table~\ref{tab1} compare the characteristics of the typical
  instrumentally smeared profile with the deconvolved
  profile. Apparently, the instrumental smearing may bias the results
  only insignificantly. Since the optimum filter removes noise, the
  bottom left subpanel of Figure~\ref{fig9} shows the differences
  $\Delta I=100(I_i - F_i)/I_i$, where $I_i$ represents the
  deconvolved data and $F_i$ is the model function. The left subpanel
  of Figure~\ref{fig9} demonstrates that the blue wing excess persists
  even after deconvolving the typical profile, because the difference
  $\Delta I$ at $\Delta\lambda \approx -0.33$\,\AA\ is
  3.3\%. Therefore we conclude that the blue wing excess is not due to
  the instrumental smearing. This might suggest inadequacy of the
  double-Gaussian model in representing the typical \hei\ profiles
  emitted by the target prominence.

  \section{Alternative Fitting Models}

\begin{figure}
    \centerline{
      \includegraphics{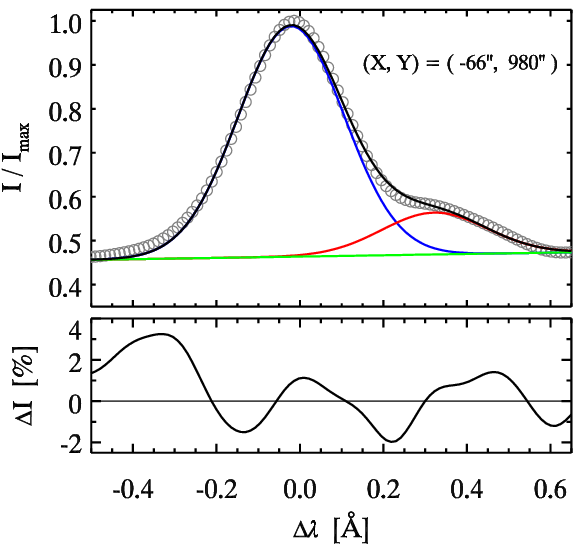}
      \includegraphics{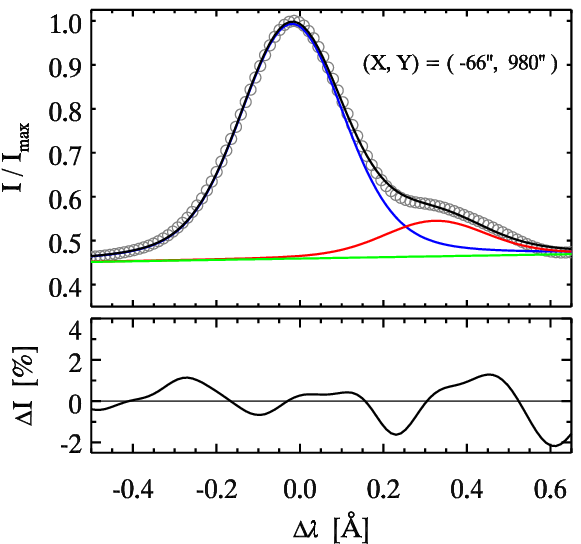}
    }
    \caption{The double-Gaussian and the double-Voigt fits ({\it black
        lines} in the {\it left and right panels}, respectively) of
      the deconvolved \hei\ Stokes~{\it I} profile ({\it gray
        circles}) shown in the left panel of Figure~\ref{fig5}. The
      blue and red fit components and the linear background are
      represented by the {\it blue, red, and green lines}. The {\it
        bottom subpanels} display the differences between the
      deconvolved \hei\ profile and the fits.}
    \label{fig9}
  \end{figure}

  In this study we adopt the double-Gaussian fitting model of the
  \hei\ profiles. An indicator that the fitting model is inadequate is
  not only the value of $\chi^2$, but also the shape of residuals. If
  a model represents data correctly, then the residuals or differences
  $\Delta I$ should be featureless with no systematic structure other
  than random excursions reflecting the noise variations. However, the
  residuals in Figure~\ref{fig5} and differences in the left subpanel
  of Figure~\ref{fig9} are not featureless, because they show
  quasi-periodic variations, which are particularly apparent in the
  blue wing of the blue component at $\Delta\lambda \approx
  -0.35$\,\AA. This may indicate an inadequacy of the adopted
  seven-parameter double-Gaussian model. The blue wing excess is also
  apparent in the \hei\ Stokes~{\it I} profiles shown in
  \citet{Landmanetal1977,HouseandSmart1982,LopezAristeandCasini2002,LopezAristeandCasini2003,Casinietal2003,Casinietal2009};
  and \citet{LopezAristeandAulanier2007}. Therefore, in the following
  we examine alternative \hei\ fitting models with the aim to propose
  a new model that appears to be characterized by small and
  featureless differences $\Delta I$ and negligible blue wing excess
  as well.

  \subsection{Double-Lorentzian Model}
  \label{lorentz}

  Recently, \citet{GonzalezManriqueetal2016} applied a
  double-Lorentzian profile in fitting blended components of the
  He\,{\sc i}~10\,830\,\AA\ triplet observed by the {\it GREGOR
    Infrared Spectrograph} (GRIS) at the 1.5-meter GREGOR solar
  telescope. Following this approach, we test the seven-parameter
  double-Lorentzian model using the typical profile shown in the left
  panel of Figure~\ref{fig5}. For this particular profile, the model
  renders much higher $\chi^2$ values and residuals than the
  seven-parameter double-Gaussian model. Moreover, it substantially
  overestimates the observed blue wing intensity of the blue component
  at $\Delta\lambda \approx -0.35$\,\AA. Hence, the double-Lorentzian
  model renders an inadequate representation of our data and is
  dismissed.

  \subsection{Double-Voigt Model}
  \label{voigt1}

  Since the double Gaussian underestimates and the double Lorentzian
  overestimates the observed blue wing intensity of the blue
  component, the double-Voigt function might render a natural
  compromise. In fact, the Voigt function has been employed by
  \citet{Elste1953} and \citet{Landmanetal1977} in fitting the
  \hei\ profiles. Therefore, we test a model consisting of two Voigt
  functions normalized to their maxima. The model has seven free
  parameters: the peak amplitudes $I_{\rm blue,\, red}$; the damping
  parameter $\varGamma$ and FWHM common for both components; the
  spectral position of the blue component peak $\lambda_{\rm blue}$;
  and two coefficients of the polynomial. The parameters of this model
  for the typical deconvolved profile are given in column 4 of
  Table~\ref{tab1}. The inferred damping parameter is shown there both
  in wavelength units as $\varGamma_\lambda$ and in frequency units as
  the decadic logarithm of $\varGamma_\nu$.  They are linked by the
  conversion formula $\varGamma_\nu = c\varGamma_\lambda/\lambda^2$,
  where $c$ is the speed of light and $\lambda$ is the wavelength of
  \hei. The right panel of Figure~\ref{fig9} shows the typical
  deconvolved \hei\ profile by the gray circles, and its double-Voigt
  fit is represented by the black line. The bottom subpanel
  illustrates that the model is free of the blue wing excess, compared
  with the left subpanel and subpanels in Figure~\ref{fig5}, and
  renders a satisfactory fit with featureless differences $\Delta I$
  smaller than 2\%. Although the model seems to comply with our
  requirements imposed on a new adequate model, we postpone a final
  judgement about its physical adequacy to Section~\ref{voigt2} of the
  Discussion.

  \begin{table}
    \caption{Parameters of fitting models shown in Figures~\ref{fig5},
      \ref{fig9}, and \ref{fig10}. Types of fitting models and
      profiles in particular columns are given below the table. Equal
      FWHM and $\varGamma$ are assumed for components of the double-Voigt
      fit. Columns 5 and 6 list the parameters of the cool and hot
      component (in parentheses) of the multi-Gaussian fit.}
    \label{tab1}
    \begin{tabular}{l r@{}l r@{}l r@{}l r@{}l r@{}l r@{}l}
      \hline
      Spectral characteristic & \multicolumn{2}{c}{\quad1} & \multicolumn{2}{c}{\quad2} & \multicolumn{2}{c}{\quad3} & \multicolumn{2}{c}{4} & \multicolumn{2}{c}{5} & \multicolumn{2}{c}{6} \\
      \hline
      FWHM$_{\rm blue}$  [ \AA\ ] & \quad0.&31 &  \quad0.&32 & \quad0.&30 & \multirow{2}{*}{0.}&\multirow{2}{*}{25} & \multirow{2}{*}{0.28}&\multirow{2}{*}{\,(0.65)} & \multirow{2}{*}{0.26}&\multirow{2}{*}{\,(0.56)} \\
      FWHM$_{\rm red}$  [ \AA\ ]  &     0.&31  &      0.&35 &      0.&30  &     &   &    &        &     & \\
      $I_{\rm blue}$/$I_{\rm max}$  &    --&    &    --&   &   --&    &  --&   & 0.94&\,(0.52)   & 0.86&\,(0.60)  \\
      $T$ [ kK ]                &   --&    &   --&   &   --&   &   --&   & 11.5&\,(91)   & 9&\,(65)  \\
      Peak intensity ratio      &      5.&5  &      4.&9  &       5.&5  &   6.&6  & 6.6&\,(8.0) &  6.3&\,(4.1)  \\
      Width ratio              &      1.&0  &      0.&9  &       1.&0  &   1.&0  & 1.0&\,(1.0) &   1.0&\,(1.0) \\
      $\varGamma_\lambda$  [ \AA\ ]/$\log(\varGamma_\nu\,[{\rm s}^{-1}])$   &    --&    &    --&  &    --&   &   0.5&/10.6  &   --&  &  --&   \\
      \hline
      \multicolumn{13}{l}{1 \quad Double-Gaussian fit of the typical smeared profile, the left panel of Figure~\ref{fig5}.} \\ 
      \multicolumn{13}{l}{2 \quad Double-Gaussian fit of the average smeared profile, the right panel of Figure~\ref{fig5}.} \\ 
      \multicolumn{13}{l}{3 \quad Double-Gaussian fit of the typical deconvolved profile, the left panel of Figure~\ref{fig9}.} \\ 
      \multicolumn{13}{l}{4 \quad Double-Voigt fit of the typical deconvolved profile, the right panel of Figure~\ref{fig9}.} \\ 
      \multicolumn{13}{l}{5 \quad Multi-Gaussian fit of the typical deconvolved profile, the left panel of Figure~\ref{fig10}.} \\ 
      \multicolumn{13}{l}{6 \quad Multi-Gaussian fit of the average deconvolved profile, the right panel of Figure~\ref{fig10}.} \\ 
    \end{tabular}
  \end{table}

  \subsection{Multi-Gaussian Model}   
  \label{multi1}

  The problem of the enhanced wing emissions in the \hei\ and He\,{\sc
    i}~10\,830\,\AA\ lines was addressed by \citet{Landmanetal1977}
  and \citet{KotrcandHeinzel1989}, who invoked two-temperature models
  of prominence structure. Inspired by this approach to explain the
  surplus of blue wing emission, we construct a multi-Gaussian model
  consisting of two double-Gaussians with different line widths
  representing the cool and hot components of a prominence. The model
  has nine free parameters: the peak intensities of the blue
  components $I_{\rm blue}$ (two parameters); the ratios of the peak
  intensities $I_{\rm blue}/I_{\rm red}$ constrained by the limit
  values 0.001 and 8 (two parameters); the FWHM common for the blue
  and red components, but different for the cool and hot ones (two
  parameters); the spectral position of the blue component peak
  $\lambda_{\rm blue}$ (one parameter); and two coefficients of the
  polynomial (two parameters). The parameters of this model for the
  typical and average deconvolved profiles are listed in columns 5 and
  6 of Table~\ref{tab1}, showing the hot component parameters in
  parentheses. It also shows the blue component intensities $I_{\rm
    blue}/I_{\rm max}$ and the kinetic temperatures $T$ of the cool
  and hot components estimated by the formula:
  \begin{equation}
    \label{eq1}
    T = \frac{m_{\rm He}}{2k}\left \{ \left(\frac{\rm FWHM}{2\sqrt{\ln{2}}}\frac{c}{\lambda}\right)^2 - v^2 \right \}\,,
  \end{equation}
  where $m_{\rm He}$ is the mass of a helium atom, $k$ is the
  Boltzmann constant, $c$ is the speed of light, $\lambda$ is the
  wavelength of \hei, and $v$ is the microturbulent velocity of
  5\,\kms\ for both components \citep[\textit{e.g.}][]{GouttebrozeHeinzelVial1993,LegerandPaletou2009}.

\begin{figure}
    \centerline{
      \includegraphics{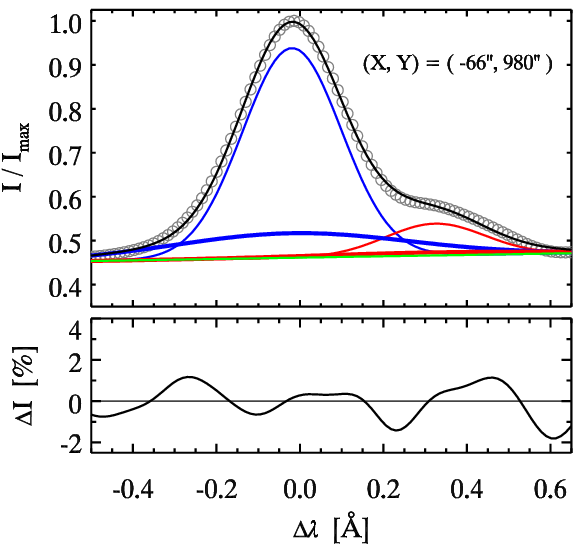}
      \includegraphics{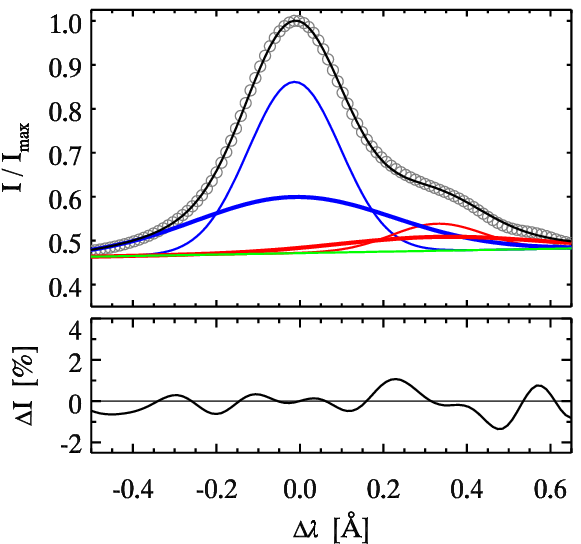}
    }
    \caption{The multi-Gaussian fits ({\it black lines}) of the
      typical \hei\ Stokes~{\it I} profile ({\it left}) and the
      average profile ({\it right}) shown by the {\it gray
        circles}. Instrumental smearing of the profiles is deconvolved
      by the optimum filter. The {\it thin blue and thin red lines}
      correspond to the cool component and the thick lines to the hot
      component. The {\it bottom subpanels} display the differences
      between the deconvolved profiles and their fits.}
    \label{fig10}
  \end{figure}

  Figure~\ref{fig10} displays the typical and average \hei\ profiles
  free of the instrumental smearing by the gray circles and their
  multi-Gaussian fits, comprising the cool (thin blue and red lines)
  and hot components (thick lines), shown by the black lines. The
  bottom subpanels illustrate that the model is free of the blue wing
  excess when compared with respective subpanels in Figures~\ref{fig5}
  and \ref{fig9}, and renders satisfactory fits with differences
  $\Delta I$ smaller than 2\%, similar to the double-Voigt
  model. Thus, the multi-Gaussian model also complies with the
  requirements we impose on our new fitting model. We postpone a final
  judgement about its physical adequacy to Section~\ref{multi2} of the
  following Discussion.

  \section{Discussion}

  \subsection{Small Stokes~{\it Q, U, V} Amplitudes}
  \label{small}

  An important fact characterizing the target prominence is the
  surprisingly small amplitude of the Stokes~{\it Q, U, V} profiles,
  typically about 0.2 -- 0.4\% of the Stokes~{\it I} peak
  intensities. The previous studies by, {\it e.g.},
  \citet{LandiDeglInnocenti1982,Querfeldetal1985,LopezAristeandCasini2002,LopezAristeandCasini2003,Casinietal2003,Casinietal2009,RamelliandBianda2005,LopezAristeandAulanier2007};
  and \citet{Paletou2008} reported the Stokes~{\it Q, U, V} amplitudes
  ranging from $4\times10^{-3}$\,\% to 2.3\,\%. Is the weak
  polarimetric signal a possible consequence of some inherent feature
  of this particular target prominence? Or is it rather due to
  non-optimal seeing, some technical aspects of the instrument setup,
  and/or an inadequate observing or reduction procedure? An inadequacy
  of the observing or reduction procedure may probably be excluded,
  because decreasing the number of modulation cycles adopted for the
  data reduction from ten to five, as used in
  \citet{Schmiederetal2013,Schmiederetal2014} and
  \citet{Levensetal2016a,Levensetal2016b}, does not bring
  improvement. A very plausible explanation was suggested by
  \citet{LopezAristeandCasini2003}. To quote them: ``It has been
  suggested by \citet{Querfeldetal1985} that line-of-sight integration
  or limited spatial resolution might result in a depression of the
  degree of linear polarization of the observed radiation, which would
  be wrongly interpreted in terms of Hanle depolarization by stronger
  fields.'' Therefore, we conclude that a combination of non-optimal
  seeing, the long integration time, and THEMIS instrumental
  characteristics very probably caused a degradation of the spatial
  resolution. This is most likely the culprit for the small
  Stokes~{\it Q, U, V} amplitudes.

  \begin{table}
    \caption{Statistics of the \hei\ spectral characteristics of the
      quiescent prominence derived from the histograms in
      Figures~\ref{fig6} and \ref{fig7}. The positive velocity
      indicates redshift of the \hei\ line center. The location of the
      second peak in the distributions of FWHM$_{\rm blue}$ and
      FWHM$_{\rm red}$ is shown in parentheses.}
    \label{tab2}
    \begin{tabular}{l r@{}l r@{}l r@{}l}
      \hline
      Spectral characteristic & \multicolumn{2}{c}{Median} & \multicolumn{2}{c}{Peak location} & \multicolumn{2}{c}{Sigma} \\ 
      \hline
      Doppler velocity  [ \kms\ ]                          &   \quad0.&3  &   \qquad0.&4   &   \quad1.&7 \\
      FWHM$_{\rm blue}$  [ \AA\ ]                       &   0.&31   &    0.25\,&(0.30)       &   --& \\
      FWHM$_{\rm red}$  [ \AA\ ]                        &    0.&29   &    0.22\,&(0.31)      &  --& \\
      Component peak intensity ratio              &    5.&4     &    5.&5      &    0.&4 \\
      Component width ratio                           &    1.&03     &    1.&04      &   0.&18 \\
      \hline
    \end{tabular}
  \end{table}

  \subsection{Spectral Characteristics of the Double-Gaussian Model}

  While Table~\ref{tab1} lists the spectral characteristics of
  the typical and average \hei\ profile inferred by three
  different fitting models, Table~\ref{tab2} summarizes
  statistics of the \hei\ spectral characteristics derived from
  the histograms in Figures~\ref{fig6} and \ref{fig7}, which
  represent the double-Gaussian model.


  The most important fact characterizing the target prominence is the
  component peak intensity ratio of $5.5\pm0.4$ (Table~\ref{tab2}),
  which differs from the optically thin limit 8. However, this is not
  a new finding. \citet{HouseandSmart1982} previously reported a peak
  ratio of $6.6\pm0.4$. In a study of eight prominences,
  \citet{Athayetal1983a} found the \hei\ peak intensity ratios ranging
  from 6.1 to 7.6. Finally, \citet{LopezAristeandCasini2002} and
  \citet{WiehrandBianda2003} reported peak intensity ratios of 6.8 and
  6, respectively. While the former two studies employed the
  double-Gaussian model, the method of inferring the ratios in the
  latter two articles is unclear. Thus, the measured peak intensity
  ratio of $5.5\pm0.4$ is on the lower limit of the values reported so
  far.


  Many of the published Doppler velocity measurements of prominences
  were made in \ha\ by the MSDP instrument \citep{Mein1977,Mein1991},
  but reports on the \hei\ velocity measurements are infrequent. A
  search of the NASA Astrophysics Data System provides us only with
  the study by \citet{Prasadetal1999}, presenting the line-of-sight
  velocity distribution over a quiescent prominence observed in
  \hei\ with a standard deviation of 1.8\,\kms. This agrees with the
  deviation of 1.7\,\kms\ derived from our measurements. The interval
  of $\pm5$\,\kms, presented in the top right panel of
  Figure~\ref{fig6}, is characteristic for the quiescent type of
  prominences and is fully in line with \hei\ \citep{Prasadetal1999},
  \ha\ \citep[\textit{e.g.}][]{Schmiederetal2010}, He\,{\sc
    i}~10\,830\,\AA\ \citep[\textit{e.g.}][]{OrozcoSuarezetal2014},
  and Mg\,{\sc ii}~k\,\&\,h \citep[\textit{e.g.}][]{Vialetal2016}
  Doppler-velocity measurements in quiescent prominences.


  The sharp lower limit of the measured width FWHM$_{\rm blue}$ of the
  dominant \hei\ blue component at about 0.23\,\AA\ (the bottom panel
  of Figure~\ref{fig6}) agrees well with the lower limit of Doppler
  widths of \hei\ reported by \citet{Athayetal1983a}. However, they
  reported the upper limit of the observed FWHM at 0.32\,\AA, while
  the histogram in the bottom panel of Figure~\ref{fig6} shows a
  continuous distribution of broad profiles beyond this
  value. \citet{HouseandSmart1982} measured a nearly equal FWHM of
  0.28\,\AA\ for both components. Similarly, the median of 15 FWHM
  values reported in \citet[][Table~1]{Landmanetal1977} was
  0.27\,\AA. Most of the \hei\ FWHMs measured by
  \citet{Prasadetal1999} were within the interval from 0.32\,\AA\ to
  0.56\,\AA, in good agreement with our results shown in the bottom
  panel of Figure~\ref{fig6}.
  

  Finally, we find width ratios of the \hei\ blue and red components
  of $1.04\pm0.18$ over almost the entire prominence. Their close
  equality was also reported by \citet{HouseandSmart1982} and
  \citet{Athayetal1983a}. The width equality is a strong confirmation
  that the \hei\ multiplet components in the plasma have a common
  origin.
  
  \subsection{Toward a New Fitting Model}

  In the following we examine in detail the physical adequacy of the
  double-Voigt model and the multi-Gaussian model as alternative
  models of \hei.

  \subsubsection{Double-Voigt Model}
  \label{voigt2}

  We have shown that the double-Voigt model of the typical profile
  requires a damping parameter $\varGamma_\lambda =
  0.5$\,\AA\ corresponding to $\log(\varGamma_\nu\,[{\rm s}^{-1}]) =
  10.6$. How does this observed value compare with the theoretical
  parameter given as the sum of the natural damping parameter
  $\varGamma_{\rm RAD}$ and the collisonal (or Van der Waals) damping
  parameter $\varGamma_{\rm VdW}$ \citep[\textit{e.g.}][]{Rutten2003}?
  The latter parameter represents collisions of the emitting helium
  atoms with the neutral hydrogen atoms. The sum of the relevant
  Einstein coefficients of spontaneous emission $A$, adopted from the
  National Institute of Standards and Technology Atomic Spectra
  Database \citep[NIST ASD,][]{Kramidaetal2016}, gives the value
  $\log(\varGamma_{\rm RAD}\,[{\rm s}^{-1}]) = 7.7$. Assuming a
  prominence gas pressure of 1\,dyn\,cm$^{-2}$ and a temperature of
  11\,kK in combination with the line-broadening theory by
  \citet{Warner1967}, one can estimate for \hei\ the value of
  $\log(\varGamma_{\rm VdW}\,[{\rm s}^{-1}]) = 4.4$. Thus, the
  observed damping parameter is about three orders of magnitudes
  larger than the theoretical parameter. Apparently, this disqualifies
  the double-Voigt model from any further consideration because it is
  physical inadequate. This conclusion is in accord with
  \citet{Landmanetal1977}, who also dismissed the physical adequacy of
  the Voigt-shaped absorption coefficient for \hei. They inferred the
  broadening parameter
  \begin{equation}
    \label{eq2} 
    a = \frac{2\sqrt{\ln{2}}}{\rm FWHM}\frac{\lambda^2}{c}\frac{\varGamma_\nu}{4\pi}
  \end{equation}
  on the order of magnitude 0.1 corresponding to
  $\log(\varGamma_\nu\,[{\rm s}^{-1}]) = 10.3$ for FWHM =
  0.27\,\AA. Because this value was also orders of magnitude
  greater than the values calculated on the basis of commonly
  accepted prominence plasma parameters, they ruled out
  collisional broadening as a realistic solution of the blue
  wing excess. This is in accord with \citet[][p.\,712 in
    Volume~II]{LandiDegl'InnocentiandLandolfi2004}, who claimed that
  collisions can be neglected in forming \hei.

  \subsubsection{Multi-Gaussian Model}
  \label{multi2}

  Focusing on the typical \hei\ profile (the left panel of
  Figure~\ref{fig10} and column 5 in Table~\ref{tab1}), we assess the
  multi-Gaussian model, which is similar to the two-temperature model
  of \hei\ applied in \citet{Landmanetal1977} and \citet{Landman1981}.
  The medians of 32 FWHM values, reported in Table~2 of
  the latter study for the cool and hot components, are about
  0.22\,\AA\ and 0.59\,\AA, respectively, and thus smaller than
  the equivalent FWHMs 0.28\,\AA\ and 0.65\,\AA\ inferred by the
  typical profile.
  Remarkably, the temperatures 11.5\,kK and 91\,kK of the cool
  and hot components of the multi-Gaussian model agree very well
  with the central and boundary temperatures of 10\,kK and
  100\,kK for the 2D model of prominence fine structure applied
  in \citet{Gunaretal2007} and \citet{Schwartzetal2015}. This
  may be evidence of the physical adequacy of the multi-Gaussian
  model.


  While the component peak intensity ratio of 8 of the hot component
  (Table~\ref{tab1}) implies its negligible optical thickness, the
  ratio of 6.6 of the cool component can be interpreted in terms of
  the prominence geometrical width and optical thickness based on the
  2D multi-thread prominence model by \citet{Leger2008}. Figures~7.21
  and 7.22 in \citet{Leger2008} present the \hei\ component peak
  intensity ratio as a function of the number of 1.2\,Mm wide
  prominence threads and their total optical thickness for the
  temperatures 8\,kK and 17\,kK. Taking the temperature 11.5\,kK of
  the cool component of the target prominence and the ratio of 6.6,
  one can estimate by interpolating the values in these figures that
  the target prominence is composed from of 14 threads with a total
  geometrical width of 17\,Mm and an optical thickness of
  0.3. Remarkably, the width, corresponding to 23\,arcsec, is
  comparable with the typical width of a filament (see the right panel
  of Figure~\ref{fig3}). However, the optical thickness of 0.3 is
  about an order of magnitude greater than the thicknesses reported in
  \citet{Landmanetal1977,Landman1981}; and \citet{Lietal2000}. In this
  context, \citet{Landmanetal1977} commented on their method and
  results by stating: ``The computed line shapes are relatively
  insensitive to the $\tau_{0i}$ because of the smallness of these
  quantities.'' For this reason, we may conclude that the prominence
  parameters inferred by the typical profile support the physical
  adequacy of the multi-Gaussian model with important implications for
  the interpretation of \hei\ spectropolarimetry by current inversion
  codes.

  \section{Summary and Conclusions}


  In this study we analyze the observations of a quiescent, tree-like
  prominence scanned by the THEMIS spectrograph on 2 August 2014 in
  the \hei\ multiplet. In a broad-band \ha\ image the distribution of
  its relative intensity with respect to the disk center has the
  median at 17\%.


  The double-Gaussian model of the \hei\ Stokes~{\it I} profiles shows
  wide distributions of the FWHM with two maxima at 0.25\,\AA\ and
  0.30\,\AA\ for the \hei\ blue component and at 0.22\,\AA\ and
  0.31\,\AA\ for the red component. The FWHM distributions have
  medians at 0.31\,\AA\ and 0.29\,\AA\ for the blue and red component,
  respectively. We find width ratios of the \hei\ components of
  $1.04\pm0.18$ over almost over the entire prominence. The width
  equality is a strong confirmation of the common origin of the
  multiplet components in the plasma. This model yields a
  \hei\ component peak intensity ratio of $5.5\pm0.4$, which differs
  from the value of 8 expected in the optically thin limit. Most of
  the measured Doppler velocities are from the interval
  $\pm5$\,\kms\ with a standard deviation of $\pm 1.7$\,\kms\ around
  the peak value of $0.4$\,\kms\ (Section~\ref{results}). The
  pixel-by-pixel comparisons of the \hei\ spectral characteristics in
  Figure~\ref{fig8} and their values in Tables~\ref{tab1} and
  \ref{tab2} may become valuable in future modeling of \hei. However,
  we have shown that the double-Gaussian model fails to reproduce the
  observed \hei\ blue wing intensities and leads to quasi-periodic
  residuals with more than one-sigma amplitudes
  (Section~\ref{Double-Gaussian-Fit}).


  We demonstrate that the blue wing excess persists even after
  correcting the typical \hei\ profile for the instrumental profile of
  the THEMIS spectrograph, suggesting the inadequacy of the
  double-Gaussian model (Section~\ref{effect}). To investigate this
  issue we test the double-Lorentzian and the double-Voigt model
  showing that the former produces unsatisfactory fits and the latter
  is physically invalid (Sections~\ref{lorentz}, \ref{voigt1}, and
  \ref{voigt2}).

  
  With the goal of identifying an adequate fitting model, we examine
  the multi-Gaussian model consisting of two double-Gaussians with
  different line widths, representing the cool and hot components of
  the prominence. This model adequately reproduces the typical
  \hei\ profile, indicating temperatures for the cool and hot
  components of about 11.5\,kK and 91\,kK, respectively. The cool and
  hot components of the typical \hei\ profile have component peak
  intensity ratios of 6.6 and 8, implying a prominence geometrical
  width of 17\,Mm and an optical thickness of 0.3 for the cool
  component, while the optical thickness of the hot component is
  negligible (Sections~\ref{multi1} and \ref{multi2}). These
  prominence parameters seem to be realistic, which supports the
  physical adequacy of the multi-Gaussian model. This has important
  implications for the interpretation of \hei\ spectropolarimetry
  using current inversion codes.


  These conclusions are based on data taken during a non-optimal
  seeing period, which is most likely the culprit for the observed
  small amplitudes of the Stokes~{\it Q, U, V} profiles, which
  typically reach only 0.2 -- 0.4\% of the Stokes~{\it I} peak
  intensities (Sections~\ref{data} and \ref{small}). A statistically
  larger sample of data, taken under more favorable seeing conditions,
  is needed to confirm these conclusions.

  \begin{acks}
    We thank an anonymous referee for valuable comments, which
    improved the article substantially. J.\,Koza is grateful to
    P.\,Heinzel and E.\,Dzif\v{c}\'{a}kov\'{a} for fruitful
    discussions, comments, and suggestions. J.\,Koza and M.\,Koz\'{a}k
    would like to thank B.\,Gelly, the director of the THEMIS solar
    telescope, and the technical team for their support during their
    THEMIS observing campaign. The authors thank M.\,Saniga for
    language corrections of the article.
    This work was supported by the Science Grant Agency project VEGA
    2/0004/16. The THEMIS observations were taken within the
    EU-7FP-SOLARNET Transnational Access and Service Programme (High
    Resolution Solar Physics Network -
    FP7-INFRAS\-TRUCTURES-2012-1). This article was created by the
    realization of the project ITMS No. 26220120029, based on the
    supporting operational Research and development program financed
    from the European Regional Development Fund. 
    This work uses GONG data obtained by the NSO Integrated Synoptic
    Program (NISP), managed by the National Solar Observatory, which is
    operated by the Association of Universities for Research in Astronomy
    (AURA), Inc. under a cooperative agreement with the National Science
    Foundation. The AIA data used here are courtesy of SDO (NASA) and the
    AIA consortium. The STEREO\,B data used here were produced
    by an international consortium of the Naval Research Laboratory (USA),
    Lockheed Martin Solar and Astrophysics Lab (USA), NASA Goddard Space
    Flight Center (USA), Rutherford Appleton Laboratory (UK), University
    of Birmingham (UK), Max-Planck-Institut for Solar System Research
    (Germany), Centre Spatiale de Liege (Belgium), Institut d'Optique
    Theorique et Appliquee (France), and Institut d'Astrophysique Spatiale
    (France).
    This research has made use of NASA Astrophysics Data System.\\ \\       
    {\noindent \bf Disclosure of Potential Conflicts of Interest} The authors
    declare that they have no conflicts of interest.
  \end{acks}

\end{article} 
\end{document}